\documentclass[12pt]{article}
\usepackage{amssymb}
\usepackage{color,graphicx}
\usepackage{amsmath}

\usepackage{cite}
\usepackage{float}

\newcommand{\p}{\bot}
\newcommand{\dd}{\partial}
\newcommand{\de}{\delta}

\newcommand{\e}{\varepsilon}
\newcommand{\ls}{\left(}
\newcommand{\rs}{\right)}

\newcommand{\La}{\Lambda}
\newcommand{\m}{\mu}
\newcommand{\n}{\nu}

\newcommand{\pa}{\scriptscriptstyle \|}
\newcommand{\al}{\alpha}
\newcommand{\ga}{\gamma}

\newcommand{\disn}[2]{$$\displaylines{\refstepcounter{equation}\label{#1}\hskip 1em minus 1em #2\hfilneg}$$}
\newcommand{\nom}{\hfil\hskip 1em minus 1em (\theequation)}
\newcommand{\ns}{\hfill\cr\hfill}

\begin{document}

\title{Pauli-Villars Regularization and Light Front  Hamiltonian in (2+1)-dimensional Yang-Mills Theory}

\author{
M.Yu.~Malyshev\thanks{E-mail: mimalysh@yandex.ru},
S.A.~Paston,
E.V.~Prokhvatilov,
R.A.~Zubov,
V.A.~Franke\\
{\it Saint Petersburg State University, St.-Petersburg, Russia}
}
\date{\vskip 15mm}
\maketitle

\begin{abstract}
The renormalization problem of (2+1)-dimensional
Yang-Mills theory quantized on the light front is considered. Extra
fields analogous to those used in Pauli-Villars regularization are
introduced to restore perturbative equivalence between such
quantized theory and conventional formulation in Lorentz
coordinates. These fields also provide necessary ultraviolet
regularization to the theory. Obtained results allow to construct renormalized Hamiltonian of the theory on the light front.
\end{abstract}


\maketitle

\section{Introduction}
The present paper is devoted to quantization of field theory on the light front. Yu.V.~Novozhilov was interested in this subject for many years, and he worked in this direction together with the part of authors of this paper.

Quantization of field theory on the light front (LF) \cite{dir,Bakker.Bassetto.Brodsky.Broniowski.Dalley.Frederico.Glazek.Hiller.Ji.Karmanov.et.al.arXiv2013} allows
to simplify the description of the
vacuum state. This makes the application of
nonperturbative Hamiltonian approach to the bound state and mass
spectrum problem more convenient \cite{paul,
Bakker.Bassetto.Brodsky.Broniowski.Dalley.Frederico.Glazek.Hiller.Ji.Karmanov.et.al.arXiv2013}. The LF can be defined by the
equation $x^+=0$ where $x^{+} = \frac{x^0 + x^1}{\sqrt{2}}$ plays
the role of time ($x^0, x^1, x^{\p}$ are Lorentz coordinates with
 $x^{\p}$ denoting the remaining spatial coordinates). The role of usual space coordinates is
 played
by the LF coordinates $x^{-} = \frac{x^0 - x^1}{\sqrt{2}}, \,
x^{\p}$.

The generator $P_-$ of translations in $x^-$ is kinematical
\cite{dir} (i.e. it is independent of the interaction and
quadratic in fields, as a momentum in a free theory). On the other
side it is nonnegative ($P_-\geqslant0$) for quantum states with
nonnegative mass squared. So the state with the minimal eigenvalue
$p_-=0$ of the momentum operator $P_-$ can describe (in the case
of the absence of the massless particles) the vacuum state, and it
is also the state minimizing $P_+$ in Lorentz invariant
theory. This means that the physical vacuum turns out to coincide with the
mathematical one. It is possible to introduce the Fock space on
this vacuum and formulate in this space the eigenvalue problem for
the operator $P_+$ (i.e. the LF Hamiltonian). In this way one
can find the spectrum of mass $m$ in the subspace with fixed
values of the momenta $p_-, p_{\bot}$ \cite{paul}:
\disn{1aa}{
P_+|p_-,p_{\bot}\rangle=\frac{m^2+p_{\bot}^2}{2p_-}|p_-,p_{\bot}\rangle,
\nom}
see details in the review paper \cite{NPPF}.

The theory on the LF has the singularity at $p_- = 0$ \cite{NPPF}.
To regularize it we use the cutoff $p_-\geqslant\e>0$.
In the present paper we consider a way to construct  the
perturbatively renormalized Hamiltonian on the LF (on difficulties
to solve this problem see
\cite{Wilson,Glazek,Glazek.Wilson.Phys.Rev.D1993,
Glazek.Wilson.Phys.Rev.D1994,
Glazek.ActaPhys.Polon.B2008,Glazek.Phys.Rev.D2001,
Glazek.Wieckowski.Phys.Rev.D2002}).
We consider this problem for (2+1)-dimensional Yang-Mills theory quantized on
the LF  where the above mentioned regularization
$|p_-|\geqslant\e>0$ is applied. Unlike the early considered case
of (3+1)-dimensional QCD \cite{tmf99}, for (2+1)-dimensional model we can
construct renormalized LF Hamiltonian  containing no new unknown
renormalization parameters.

It was shown in papers \cite{burlang, tmf97,NPPF} that some
diagrams of the perturbation theory, generated by the LF
Hamiltonian, and corresponding diagrams of the conventional
perturbation theory in Lorentz coordinates can differ.
 It was found that one can overcome this difficulty by addition of new (in particular, nonlocal) terms to the
canonical LF Hamiltonian \cite{tmf99,tmf02,Yad.Fiz.2005}. In the case of gauge field theory the infinite number of such terms appears \cite{tmf99,NPPF}. However one can avoide the differences between diagrams, generated by the LF
Hamiltonian, and corresponding diagrams of the conventional
perturbation theory if one adds extra ghost fields
analogous to that in the Pauli-Villars (PV) regularization \cite{tmf99,NPPF}. In this way one can avoide the infinite number of terms mentioned above. It was
also shown how to renormalize this theory ensuring the perturbative equivalence with the dimensionally regularized theory in Lorentz coordinates and the restoration of gauge
invariance. For example, in (3+1)-dimensional Quantum
Chromodynamics \cite{tmf99} this can require the inclusion of ten counterterms, necessary for UV renormalization, into
LF Hamiltonian.
It was shown that there must be the values of coefficients before these counterterms at which the
restoration of gauge invariance occurs. However one cannot
find these coefficients explicitly because of infinite number of
divergent diagrams in (3+1)-dimensional theory. So one has to
consider them as new unknown parameters.

The (2+1)-dimensional Yang-Mills
theory, considered in the present paper, is superrenormalizable, so that all renormalizing
counterterms can be found exactly via calculation of the finite number of
diagrams. This allows to carry out the renormalization of the
theory on the LF in such a way that no unknown quantities, besides the original parameters, appear. We regularize perturbative
infrared (IR) divergences introducing topological Chern-Simons (CS) term
\cite{Deser.Jackiw.Templeton.Annals.of.Physics.2000,
Deser.Jackiw.Templeton.Phys.Rev.Lett.1982}.

We analyse the perturbation theory, generated by quantization on
the LF, and investigate its equivalence to the usual covariant
perturbation theory in Lorentz coordinates. To do this we apply the method of
paper \cite{tmf99}. We use the analog of the PV regularization
to remove both UV divergences and differences between diagrams of
perturbation theory on the LF and corresponding diagrams of covariant
perturbation theory in Lorentz coordinates. We show how to restore
gauge symmetry in the limit that removes PV regularization at the correct
renormalization of the theory.
In this way we can construct renormalized Hamiltonian
on the LF which can be used for nonperturbative calculation of mass
spectrum in accordance with (\ref{1aa}).

\section{Divergences of (2+1)-dimensional Yang-Mills theory with Chern-Simons term}
To construct the renormalized  LF Hamiltonian we have to
analyse diagrams of perturbation theory. These diagrams must
be well defined, i.e. to be free of divergences. Yang-Mills theory in (2+1)-dimensions
contains, besides usual UV divergences, also
the IR ones \cite{Deser.Jackiw.Templeton.Annals.of.Physics.2000,
Deser.Jackiw.Templeton.Phys.Rev.Lett.1982}. That makes impossible
the analysis of perturbation theory. As a solution to this problem
we introduce the CS term  \cite{Deser.Jackiw.Templeton.Annals.of.Physics.2000,
Deser.Jackiw.Templeton.Phys.Rev.Lett.1982}, generating gluon
field mass. In result we investigate the theory with the following
Lagrangian density:
\disn{2}{
{\cal L} =- \frac{1}{4} F_{\m\n}^a F^{a\m\n} + \frac{m}{2} \e^{\m\n\alpha} \ls A_{\m}^a \partial_{\n} A_{\alpha}^a  +
\frac{2}{3} g f^{abc}A_{\m}^a A_{\n}^b A_{\alpha}^c \rs,
\nom}
where $A_{\m}^a(x)$ are gluon fields
corresponding to gauge symmetry group SU(N),
$F_{\m\n}^a=\dd_{\m}A_{\n}^a - \dd_{\n}
A_{\m}^a+gf_{abc}A_{\m}^bA_{\n}^c$, $a=1,...,N^2-1$ are indices of
adjoint representation, $m$ and $g$ are parameters,
$\e^{\m\n\alpha}$ is Levy-Civita symbol.

For the construction of LF Hamiltonian we take the gauge $A_-=A^+=0$.
Its use in the action leads to the Lagrangian density in which the
contribution from the first term in (\ref{2}), having power four in
fields,  and the contribution from the CS term, having
power three in fields, disappear:
\disn{3}{
{\cal L} =- \frac{1}{4} f^{a\m\n}f_{\m\n}^a+g f^{abc}A_{+}^a A_{\p}^b \dd^+ A^{c\p}
 + \frac{m}{2} \e^{\m\n\alpha} A_{\m}^a \partial_{\n} A_{\alpha}^a,
\nom}
 where $f_{\m\n}^a=\dd_\m A^a_\n-\dd_\n A^a_\m$.
As a result, the propagator, in which the remaining part of
CS term contributes, takes in the momentum space the
following form:
\disn{4}{
\Delta_{\m\n}^{ab}=
\frac{-i\delta^{ab}}{k^2-m^2+i0}
\ls g_{\m\n}-\frac{k_{\m}n_{\n}+n_{\m}k_{\n}+
i\,m\,\e_{\m\n\alpha}n^{\alpha}}{k_{\pa}^2+i0}2k_+\rs,
\nom}
where $k_{\pa}^2=2k_+k_-$ and $n_\n$ is  lightlike  vector
with components $n_+=1$, $n_-=n_{\p}=0$. As one can see, the parameter
$m$ plays the role of the field $A_{\m}$ mass. For the
regularization of singularity at $k_-=0$ in the propagator
(\ref{4}) the Mandelstam-Leibbrandt prescription is used
\cite{Mandelstam, Leibbrandt}. Such a prescription allows to do
the Wick rotation to Euclidean momentum space for diagrams where
it is possible to analyse UV divergences of Feynman diagrams in the
standard way.

The interaction term in the equation (\ref{3}) leads to the vertex which contains derivative with upper index +.
Let us remark that due to the global $SU(N)$ symmetry (note that
local, i.e. gauge symmetry, is  broken by UV regularization in our
approach) all diagrams with single external line must be equal to
zero as they are vectors in the color space.

\begin{figure}[h!]
  \centering
  \includegraphics[width=150mm]{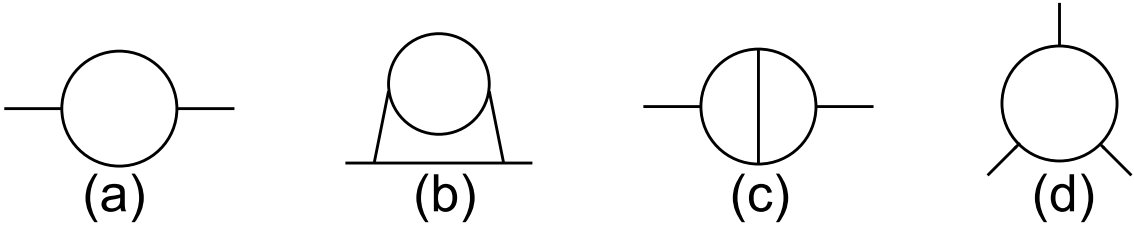}
  \caption{Divergent diagrams.}
\label{diagrams}
\end{figure}

Let us find all UV divergent Feynman diagrams
that must be renormalized. To do this we use the standard method of
estimation of the UV divergency index of Feynman integrals in
Euclidean space. In result the divergent diagrams are those shown in Fig.~\ref{diagrams}. As expected, their number is
finite.
Due to violation of Lorentz invariance by the introduction of the
 $A_-=0$ gauge the number of divergent diagrams can increase, in
 principle. To check this it is necessary to analyse not only the
 total divergency index (in all components of the momentum) but
 also the UV divergency indices corresponding to only  some part of components of the
 momentum.
Taking into account the structure of the propagator (\ref{4}) and
 the vertex one can see that only the divergency
index in transverse component $k_{\p}$ can, in principle, exceed
the total divergency index. However it is not difficult to verify
that  this case, in fact, does not realize, and the diagrams,
shown in Fig.~\ref{diagrams}, exhaust all cases of the UV
divergency.
With respect to the UV divergency index the diagram in  Fig.~\ref{diagrams}(a) is linearly
divergent  and the other diagrams are logarithmically divergent.

\section{Regularization of the theory}
\label{regulariz}
\begin{figure}[h!]
  \centering
  \includegraphics[width=164mm]{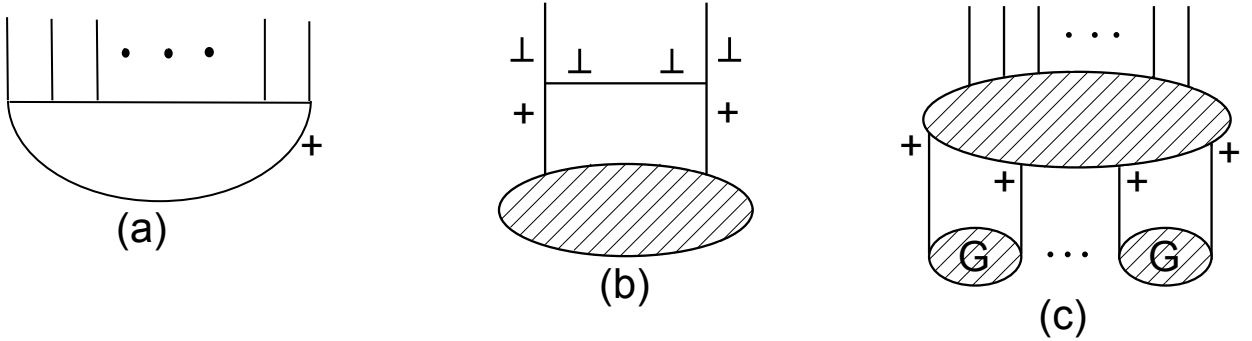}
  \caption{(a) The example of diagram which has different values in perturbation
theory on the LF and usual covariant perturbation theory.
(b) General form of diagram having mentioned above different values for the theory with single PV field.
(c) General form of diagram which can contain IR divergency.
Symbols +, $\p$ denote indices of propagators, hatched domains in diagrams denote arbitrary subdiagrams.}
\label{diagrams2}
\end{figure}

Let us assume that some UV regularization
 of the theory is introduced so that all diagrams are UV finite. As  mentioned in
the Introduction, the results of calculations of diagrams in LF
perturbation theory and usual covariant perturbation theory in
Lorentz coordinates can differ. To find these differences one can
apply the method of \cite{tmf97,tmf99, NPPF} if the regularization
$|k_-|\geqslant\e>0$ is used.
As was noted in \cite{tmf99} for the theory with the propagator containing additional pole in $k_-$ (like in eq. (\ref{4})),
these differences arise for diagrams with any number of external
lines shown in Fig.~\ref{diagrams2}(a).
As the compensation of these differences could require the addition
 of infinite number of counterterms to the action, we need some modification
of the propagator that removes these differences.
A way to do this simultaneously with the introduction of UV
regularization was proposed in \cite{tmf99}, and we will use its
analog.

The main idea of this method is  the introduction of  gauge field analog of PV
ghost fields with the simultaneous introduction of higher
(noncovariant) derivatives (that breaks gauge invariance). Let us
note that, owing to the coordinates of the light front, an
introduction of higher derivative in the form of the power one of
$\dd^2$ does not lead to a complication of canonical formalism on
the LF. This is because the action can be
transformed by integration by parts to the form containing no higher than the first derivatives
in $x^+$.
So we limit ourselves by just those higher derivatives. Let us
choose the Lagrangian density in the form
\disn{11}{
{\cal L}=\sum_{j=0}^2\ls -\frac{1}{4} f_j^{a\m\n}\ls\frac{M_j^2+\dd_{\pa}^2}{B_j}\rs f_{j,\m\n}^a+
\frac{m}{2} \e^{\m\n\alpha} A_{j,\m}^a\ls\frac{M_j^2+\dd_{\pa}^2}{B_j}\rs\partial_\n A_{j,\alpha}^a\rs+\ns+
g f^{abc}A_{\m}^a A_{\n}^b \dd^\m A^{c\n}.
\nom}
Here $f_{j,\m\n}^a=\dd_\m A_{j,\n}^a-\dd_\n A_{j,\m}^a$, the
quantity $A_{0,\m}^a$ is the physical gluon field, and
$A_{1,\m}^a$ and $A_{2,\m}^a$ are extra fields, PV fields analog.
As one can see, the quadratic part of the Lagrangian is diagonal in
these fields, and only their sum
$A_{\m}^a=A_{0,\m}^a+A_{1,\m}^a+A_{2,\m}^a$ enters into the
interaction term.
The conditions $A_{j,-}^a=0$ for physical and extra fields, analogous to the LF gauge,  are proposed.
Let us remark that the parameter $m$, giving the mass to the gauge
field, is common for all three fields $A_{j,\m}^a$, and the
differences between them are related to the values of parameters
$M_j$, $B_j$ (for conventional PV fields parameters analogous to $M_j$ correspond to masses of fields).

As the interaction contains  the sum of  fields $A_{\m}^a$,
propagators of three fields sum up in diagrams, and Feynman
integrals can be written in terms of summarized propagator
$\Delta^{ab}_{\m\n}$, i.e. the sum of individual field
propagators
\disn{11.1}{
\Delta^{ab}_{j,\m\n}=
\frac{i\delta^{ab}B_j}{\ls k^2-m^2+i0\rs\ls k_{\pa}^2-M_j^2+i0\rs}
\ls g_{\m\n}-\frac{k_{\m}n_{\n}+n_{\m}k_{\n}+
i\,m\,\e_{\m\n\alpha}n^{\alpha}}{k_{\pa}^2+i0}2k_+\rs.
\nom}
We relate the parameters $M_0$, $M_1$ and $M_2\equiv\m$ to
regularization parameters and choose the quantities $B_j$ so that
to assure the decrease of summarized propagator as
$1/k_{\pa}^6$ (see the discussion of the necessity of such
exceeded requirement in the next Sect.). On the other hand, we choose them so that
to cancel the  additional pole in $k_-$, present in the
propagator.
This can be done if one takes
\disn{13}{
B_0=\frac{M_0^4M_1^2}{\ls M_1^2-M_0^2\rs\ls M_0^2-\m^2\rs},\quad
B_1=-\frac{M_0^2M_1^4}{\ls M_1^2-M_0^2\rs\ls M_1^2-\m^2\rs},\ns
B_2=-\frac{M_0^2M_1^2\m^2}{\ls M_0^2-\m^2\rs\ls M_1^2-\m^2\rs},
\nom}
resulting in the following form of the regularized summarized
propagator:
\newpage
\disn{14}{
\Delta_{\m\n}^{ab}=
\dfrac{-i\delta^{ab}}{\ls k^2-m^2+i0\rs
\ls 1-\dfrac{k_{\pa}^2+i0}{M_0^2}\rs \ls 1-\dfrac{k_{\pa}^2+i0}{M_1^2}\rs}\times\ns\times
\frac{g_{\m\n}k_{\pa}^2-(k_{\m}n_{\n}+k_{\n}n_{\m}+
i\,m\,\e_{\m\n\alpha}n^{\alpha})2k_+}{k_{\pa}^2-\m^2+i0}.
\nom}
It is easy to check that all diagrams of Fig.~\ref{diagrams} become
finite with that propagator. After removing of the regularization
the propagator (\ref{14}) must turn into  the propagator of
nonregularized theory (\ref{4}). It is easy to see that
the following conditions must be fulfilled  when one removes the
regularization:
\disn{14.1}{
\m\to 0,\qquad
M_0\to\infty,\qquad
M_1\to\infty,\qquad
\frac{M_1}{M_0}\to\infty.
\nom}
The latter one is necessary as we want that only physical field
$A^a_{0,\m}$  remains in the regularization removing limit,
because the propagator (\ref{11.1}) of this field turns under that
condition into the expression (\ref{4}), and propagators of the
other two fields tend to zero.

\section{Comparison of perturbation theory on the LF and covariant one in Lorentz
coordinates}
Using the already mentioned above method proposed in \cite{tmf97}
it is possible to analyse the difference between the results of
calculations of diagrams in LF perturbation theory and the usual
covariant perturbation theory in Lorentz coordinates when the
regularization $|k_-|\geqslant\e$ is applied. It is possible to
calculate this difference as the difference between the diagram
for which the integration is over all momenta $k_-$ and the same
diagram for which that integration is only over the domain
$|k_-|\geqslant\e>0$ (here $k_-$  is the propagator momentum).
Thus the difference is the sum of all copies of the diagram, in
which one integrates  over the momentum $k_-$ over the domain
$|k_-|\geqslant\e>0$ at least for one of propagator momenta.
The idea of the method is the following. If one
makes for each loop momentum $k$ (which can be always identified with some propagator
momentum) the change $k_-\to k_- \e$, $k_+\to k_+/\e$, an essential dependence on $\e$ in
the integration region disappears, and one can investigate the behavior of the integrand
 for an arbitrarily complicated diagram.

In result, the above mentioned difference for an arbitrary diagram
 can be estimated in every contribution to it in the form of
 $\e^\sigma$ where $\sigma$ is determined by topology of the
 diagram, its Lorentz structure and general properties of the
 theory such as spin of the field and UV properties of the
 propagators (see details in \cite{tmf97}).

For example let us consider the diagram shown in Fig.~\ref{diagrams}(a). Let
$p_{\m}$ denotes external momentum and $k_{\m}$ denotes the loop
momentum coinciding with one of propagator momenta. In
perturbation theory on the LF this momentum is limited by the
condition $|k_-|\geqslant\e$ and, when one of the differences for
that diagram is calculated, it is limited by the condition
$|k_-|\leqslant\e$. Let us write such contribution to the
difference, when the above mentioned momentum $k$ corresponds to
the component $\Delta_{++}$ of the propagator:
\disn{17}{
\int_{-\infty}^{\infty} dk_{\p}\int_{-\infty}^{\infty}dk_+\int_{-\e}^{\e} dk_- \frac{(2p_--k_-)^2(2k_+)^2}{\ls k^2-m^2+i0\rs \ls k_{\parallel}^2-\m^2+i0\rs}\times\ns\times
\frac{2(p_+-k_+)(p_--k_-) M_0^4M_1^4}{\ls k_{\parallel}^2-M_0^2+i0\rs\ls k_{\parallel}^2-M_1^2+i0\rs\ls (p-k)^2-m^2+i0\rs}
\times\ns\times
\frac{1}{\ls2(p_+-k_+)(p_--k_-)-\m^2+i0\rs
\ls2(p_+-k_+)(p_--k_-)-M_0^2+i0\rs}
\times\ns\times
\frac{1}{\ls2(p_+-k_+)(p_--k_-)-M_1^2+i0\rs}.
\nom}
Here the first factor $(2p_--k_-)^2$ in the numerator of the
integrand  corresponds to vertices of the diagram. After the
change $k_-\to k_- \e$, $k_+\to k_+/\e$ this integral takes the
following form:
\disn{9}{
\int_{-\infty}^{\infty} dk_{\p}\int_{-\infty}^{\infty}dk_+\int_{-1}^{1} dk_- \frac{(2p_--\e k_-)^2(2k_+)^2}{\e^2\ls k^2-m^2+i0\rs \ls k_{\parallel}^2-\m^2+i0\rs}\times\ns\times
\frac{2(p_+-k_+/\e)(p_--\e k_-) M_0^4M_1^4}{\ls k_{\parallel}^2-M_0^2+i0\rs\ls k_{\parallel}^2-M_1^2+i0\rs\ls 2(p_+-k_+/\e)(p_--\e k_-)-(p_{\p}-k_{\p})^2-m^2+i0\rs}
\times\ns\times
\frac{1}{\ls2(p_+-k_+/\e)(p_--\e k_-)-\m^2+i0\rs
\ls2(p_+-k_+/\e)(p_--\e k_-)-M_0^2+i0\rs}
\times\ns\times
\frac{1}{\ls2(p_+-k_+/\e)(p_--\e k_-)-M_1^2+i0\rs}.
\nom}
In the limit $\e\to0$ this integral equals to the following
expression:
\disn{18}{
\int_{-\infty}^{\infty} dk_{\p}\int_{-\infty}^{\infty}dk_+\int_{-1}^{1} dk_- \frac{\e(2p_-)^2(2k_+)^2}{\ls k^2-m^2+i0\rs \ls k_{\parallel}^2-\m^2+i0\rs}\times\ns\times
\frac{2(-k_+)p_- M_0^4M_1^4}{\ls k_{\parallel}^2-M_0^2+i0\rs\ls k_{\parallel}^2-M_1^2+i0\rs}
\frac{1}{\ls2(-k_+)p_-+i0\rs^4}=\ns=
-2\e M_0^4M_1^4\int_{-\infty}^{\infty} dk_{\p}\int_{-\infty}^{\infty}dk_+\int_{-1}^{1} dk_-\frac{1}{(k_+p_--i0)\ls k^2-m^2+i0\rs}\times\ns\times
\frac{1}{\ls k_{\pa}^2-\m^2+i0\rs\ls k_{\pa}^2-M_0^2+i0\rs\ls k_{\pa}^2-M_1^2+i0\rs}.
\nom}
This expression tends to zero in the limit $\e\to0$ at fixed
parameters $\m,M_{0,1}$. This determines the order of the
regularizations removing: firstly $\e\to0$, then $\m\to0$,
$M_{0,1}\to\infty$, taking into account  (\ref{14.1}).

We have shown how one
of contributions to the difference disappears in the limit
$\e\to0$ for the diagram in Fig.~\ref{diagrams}(a). Following the method proposed in \cite{tmf97} one can
succeed in showing that the same is true for all possible
contributions to the differences for any diagrams of the
considered theory in any order of perturbation theory. Let us remark that in considered theory the
diagrams with all external lines, joined to single vertex, are absent (diagrams with one external line are equal to zero due to the global SU(N) symmetry and the 1--particle irreducible diagrams with two external lines, joined to single vertex, are absent due to the absence of vertices with four lines).
If such diagrams existed in the theory they could give a difference between LF perturbation theory and usual covariant perturbation theory (in
calculations on the LF such diagrams are equal to zero but they
can be nonzero in usual covariant perturbation theory, see
~\cite{tmf97}).

It may be remarked that the ultimate absence of differences for
all diagrams is owing to  sufficiently fast decrease of  the
propagator (as $1/k_{\pa}^6$). It is easy to note that it would
be sufficient to have the decrease as $1/k_{\pa}^4$ for the UV
finiteness. This can be done by introducing not two PV fields (as
have been made in Sect.~\ref{regulariz}) but only a single one.
That could simplify the theory.
However in this case the differences between calculation on the LF
and in the usual covariant perturbation theory in the limit $\e\to0$
 disappears not for all diagrams. The finite in that limit
 differences would be nonzero for infinite number of diagrams
 having the form shown in Fig.~\ref{diagrams}(b). These differences turn out to be divergent in the UV regularization
 removing limit.
To compensate them it would be necessary to add to
the LF Hamiltonian some new (having the gluon mass form) counterterm
with UV divergent coefficient (being the sum of contributions of
infinite number of differences).
We note here that one can fully avoid the appearance of unknown coefficients
before the counterterms at UV renormalization in (2+1)-dimensions (see
below). So it is reasonable to choose a variant of the theory in which
they do not appear also due to  comparison of perturbation theory
on the LF and the usual covariant perturbation theory. That is what we
do in the present paper.

\section{Analysis of longitudinal IR divergences}
\label{IK}
It was  shown in the previous section that, if we use the introduced above analog of PV regularization, the diagrams of perturbation
theory on the LF transform into the diagrams of the usual covariant
perturbation theory in the limit $\e\to0$. In the next section we show that these diagrams can be renormalized in such a way that they coincide with corresponding diagrams in dimensional regularization (and renormalization) in the regularization removing limit (\ref{14.1}). Furthermore it is possible to go to the
Euclidean form of the theory by Wick rotation because with the Mandelstam-Leibbrandt prescription the structure
of poles allows to do
that. After the Wick rotation propagator (\ref{14}) takes the form
\disn{14.2}{
\Delta_{\m\n}^{ab}=
\dfrac{-i\delta^{ab}}{\ls k^2+m^2\rs
\ls 1+\dfrac{k_{\pa}^2}{M_0^2}\rs \ls 1+\dfrac{k_{\pa}^2}{M_1^2}\rs}\times\ns\times
\frac{\de_{\m\n}k_{\pa}^2-(k_{\m}n_{\n}+n_{\m}k_{\n}-
m\,\e_{\m\n\alpha}n_{\alpha})2k_\beta n^*_\beta}{k_{\pa}^2+\m^2}.
\nom}
Here the vector $n_\m$ becomes complex vector with components
$n_0=-\frac{i}{\sqrt{2}}$, $n_1=\frac{1}{\sqrt{2}}$, $n_\p=0$, and
the vector $n^*_\beta$ is the result of its complex conjugation.
Let us note that, despite of the transition to Euclidean space, it
is possible to use the indices $-$ and $+$ as before implying by
them the contraction with vectors $n_\m$ and $n^*_\m$,
respectively. Taking into account the decomposition
$\de_{\m\n}=n_\m n^*_\n+n^*_\m n_\n+\de_{\m\p}\de_{\n\p}$ it follows that in Euclidean space one can write $a_\m
b_\m=a_+ b_- + a_- b_+ + a_{\p} b_{\p}$.

Further we analyse the limit $\m\to0$ for arbitrary diagram. As it
is seen from the form of the propagator (\ref{14.2}) the essential
(i.e. appearing at any values of external momenta)
 IR  divergences can appear in this limit at the points of the
 momentum space at which the quantities $k_{\pa}^2=k_1^2+k_2^2$
 become equal to zero for several propagator momenta
 simultaneously. Note that
every propagator gives the pole of the first order in $k_{\pa}$, and only for the component $\Delta_{+\p}$ (let us also note that
$\Delta_{-\n}=0$  and that here and further we discard color
indices).  In the paper \cite{tmf99} the analysis was carried out
of the possibility of the appearance of the longitudinal IR
divergency for the (3+1)-dimensional QCD with the analogous
regularization, when the gluon propagator  has the same
properties.
Repeating this analysis for the now considered model it is possible
to find that the above mentioned  divergence can be only
logarithmic, and it can appear only for diagrams of the type shown
in Fig.~\ref{diagrams2}(c), and only for contributions of the form
\disn{14.3}{
\Delta_{+\n}G_{\n\ga}\Delta_{\ga +}=n^*_\m\Delta_{\m\n}G_{\n\ga}\Delta_{\ga\de}n^*_\de,
\nom}
where
$G_{\n\ga}$ is one of marked in Fig.~\ref{diagrams2}(c)
subdiagrams with two external lines (not necessarily 1-particle irreducible in
general).

Let us analyse the contribution of the expression (\ref{14.3})
which gives the longitudinal IR divergence. First we write down
the contribution of  the pole in $k_{\pa}$ for one of the
quantities $\Delta_{\m\n}$ entering into (\ref{14.3}) keeping only
essential terms in which the cancelation of the pole does not take
place, and also discarding nonessential total factor:
\disn{14.4}{
n^*_\m
\frac{\de_{\m\n}k_{\pa}^2-(k_{\m}n_{\n}+n_{\m}k_{\n}-
m\,\e_{\m\n\alpha}n_{\alpha})2k_\beta n^*_\beta}{k_{\pa}^2}
G_{\n\ga}\,\,\,\,\rightarrow\,\,\,\,
-\frac{2k_\beta n^*_\beta}{k_{\pa}^2}
(k_{\n}-m\,n^*_\m\e_{\m\n\alpha}n_{\alpha})
G_{\n\ga}.
\nom}
Let us note that the vector $n^*_\m\e_{\m\n\al}n_{\al}$ has only
transversal component. At the analysis of IR divergency we can
suppose that $k_\p\ne0$
(the integration over all momenta
at $k_{\p}=0$ and $k_{\pa}=0$ does not lead to IR divergency while we have logarithmic IR divergency in $k_{\pa}$).
Then the mentioned above constant vector can be written in the
form
\disn{14.5}{
n^*_\m\e_{\m\n\al}n_{\al}=-i\frac{k_\n-k_0\de_{\n 0}-k_1\de_{\n 1}}{k_{\p}}.
\nom}
After that the essential part (\ref{14.4}) can be written, again
discarding the terms in which the cancelation of the pole takes
place, in the form
\disn{14.6}{
-\frac{2k_\beta n^*_\beta}{k_{\pa}^2}
\ls 1+i\frac{m}{k_\p}\rs k_{\n}
G_{\n\ga}.
\nom}

If gauge invariance was conserved this expression would be equal to zero
as a consequence of the Ward identities, and hence the essential
longitudinal divergences would be absent in this case. Exactly the same
result was obtained in \cite{tmf99} for (3+1)-dimensional QCD. Thus the result does not change when we take into account the
influence of the CS term.
However the used regularization violates gauge invariance,
regularizing the emerging  divergence by the parameter
$\m$. To avoid the divergence in the limit $\m\to0$ it is
necessary that simultaneously with taking this limit the UV
regularization be removed (i.e. all limits be taken simultaneously
(\ref{14.1})) and renormalizing counterterms be chosen so that in
the regularization removing limit Ward identities be satisfied
(the idea of that mechanism was supposed in \cite{tmf99}).
With the choice of counterterms in such a way that the values of
diagrams differ from renormalized results, obtained via
dimensional regularization, by the amount of the order
$\frac{1}{M_0}$ (for UV finite diagrams this is automatically true
if the product $\m M_0$ is bounded from above) the contribution of
the quantity (\ref{14.6})) can be estimated as
$O\ls\frac{1}{M_0}\rs$.
Therefore the corresponding to it total contribution to the
diagram (which is equal to zero in dimensional regularization) can
be estimated as $\frac{(\ln\m)^N}{M_0}$ (where  $N$ is the number
of subdiagrams $G$ in Fig.~\ref{diagrams2}(c)).
Let us require
that this relation tends to zero for any $N$. Then diagrams for
which the longitudinal divergency can appear will not give, in the
regularization removing limit, differences between results of
calculations in the scheme used here and those in the dimensional
regularization scheme.

So one can conclude that if one takes, for example,
\disn{14.7}{
\m\sim\frac{1}{\La},\qquad
M_0\sim\La,\qquad
M_1\sim\La^2,
\nom}
all required conditions (including the
conditions (\ref{14.1})) are satisfied in the limit $\La\to\infty$,
and one can take $\m=0$ in diagrams for the analysis of the UV divergences.

\section{Renormalization of the theory}
In the considered theory we have to renormalize only finite number
of diagrams shown in Fig.~\ref{diagrams}. We can calculate their values in the used
regularization and therefore find explicitly the coefficients of the
counterterms. This provides the coincidence of the values for
these diagrams with the results of their renormalization obtained in the
dimensional regularization. In the result, inspite of the violation of
Lorentz and gauge symmetries in the used regularization, these
symmetries are restored for the renormalized theory in the
regularization removing limit $\La\to\infty$.
Further we choose the counterterms of our theory so that the
renormalized diagrams in our regularization coincide with
renormalized diagrams in dimensional regularization.

Let us consider the diagram shown in Fig.~\ref{diagrams}(a).
After the regularization for that diagram we consider its Taylor decomposition
	in the external momentum $p_{\m}$ in the vicinity of the point $p_{\m}=0$.
Using dimensional analysis of its UV divergent parts one can find that for the linearly	divergent
	diagram it is sufficient to renormalize only the first two terms in this decomposition,
	and only the first term for the logarithmically divergent diagrams.
In the used regularization this diagram contains integrals that equal zero when the external
	upper indices are ++, +$\p$ and $\p$+.  One of two reasons can explain this.
The first one is the odd parity of integrand with respect to the one of momentum
	components.
The second reason is a possibility to express the integrand as the difference of two parts that
	cancel each other due to the symmetry under the interchange of longitudinal components
	of the integration momentum.
Note that we don't consider amputated diagrams with the upper index $-$, because in the $A_-=0$ gauge
	they don't contribute to corresponding Green functions due to contractions with propagators.
For the indices $\p\p$  we have the Euclidean form of the integral (for $p_{\m}=0$)
\disn{19}{
\int_{-\infty}^{\infty} dk_{\p}\int_{-\infty}^{\infty}dk_0\int_{-\infty}^{\infty} dk_1  \frac{(k_{\p}^2-k_0^2-k_1^2)}{(k^2+m^2)^2} R(k_0,k_1,M_0,M_1),\ns R(k_0,k_1,M_0,M_1)=\frac{M_0^4M_1^4}{\ls k_{\parallel}^2+M_0^2\rs^2 \ls k_{\parallel}^2+M_1^2\rs^2}.
\nom}
In this integral we set $\m=0$, because it is IR-finite.
Using cylindrical coordinates ($\varphi$, $\rho=\sqrt{k_0^2+k_1^2}$, $k_{\p}$) one can
	perform the integration over the angle variable that gives the factor $2\pi$:
\disn{20}{
\pi\int_0^{\infty}d\rho \int_{-\infty}^{\infty}dk_{\p} \frac{(k_{\p}^2-\rho)R(\rho,M_0,M_1)}{(\rho+k_{\p}^2+m^2)^2}= \ns
= \frac{\pi^2}{2}\int_0^{\infty} d\rho \ls \frac{1}{(\rho+m^2)^{\frac{1}{2}}}- \frac{\rho}{(\rho+m^2)^{\frac{3}{2}}} \rs R(\rho,M_0,M_1) = \ns= \frac{\pi^2m^2}{2}\int_0^{\infty}d\rho  \frac{ R(\rho,M_0,M_1)}{(\rho+m^2)^{\frac{3}{2}}} .
\nom}
Now we can remove the regularization ($M_{0,1}\to\infty$ and correspondingly $R\to1$) and
	compute the integral
\disn{21}{
\frac{\pi^2m^2}{2}\int_{m^2}^{\infty} \frac{d\rho}{(\rho+m^2)^{\frac{3}{2}}}=
\frac{\pi^2m^2}{2}\int_{m^2}^{\infty} \frac{d\rho}{\rho^{\frac{3}{2}}}=\pi^2m.
\nom}
Thus one can see that the linear divergence, in fact, is absent.
We get the same answer using dimensional regularization.
We have computed the second term in the Taylor series with an analytical computer program and found
	that this term equals zero for all concerned external indices (++, +$\p$, $\p$+ and $\p\p$)
	of the diagram.
This means that the diagram in Fig.~\ref{diagrams}(a) needs no renormalization.
Analogously, using an analytical computer program we found that the UV divergent part
	of the diagram shown in Fig.~\ref{diagrams}(d) equals zero in both regularizations and for
	all concerned external indices (+++, ++$\p$, \ldots).
We have not found an analytical answer for the divergent in the limit $M_{0,1}\to\infty$ parts
	of the remaining two diagrams.
However, these divergent parts can be calculated numerically and this is sufficient for possible
	non-perturbative computations involving the LF Hamiltonian.

In that way we have demonstrated the possibility to exactly find the counterterms that are
	needed for renormalization.
Thus now we can construct the renormalized Hamiltonian \cite{tmf99, NPPF} and use it
	for non-perturbative computations.

{\bf Acknowledgments.}
The authors thank the organizers of the International Conference
	"In Search of Fundamental Symmetries"	2-5 December 2014 in Saint Petersburg.
The authors M.Yu. Malyshev, E.V. Prokhvatilov and R.A. Zubov acknowledge
	Saint-Petersburg State University for a research grant 11.38.189.2014.


\begin{thebibliography}{10}
\newcommand{\enquote}[1]{``#1''}
\providecommand{\url}[1]{\texttt{#1}}
\providecommand{\urlprefix}{URL }
\providecommand{\eprint}[2][]{\url{#2}}

\bibitem{dir}
P.~A.~M. Dirac, \emph{Rev.~Mod.~Phys.}, \textbf{21}:3 (1949), 392--398.

\bibitem{Bakker.Bassetto.Brodsky.Broniowski.Dalley.Frederico.Glazek.Hiller.Ji.%
Karmanov.et.al.arXiv2013}
B.~L.~G. Bakker at~al., Light-Front Quantum Chromodynamics: A
  framework for the analysis of hadron physics, 2013, arXiv:1309.6333
  [hep-ph].

\bibitem{paul}
S.~J. Brodsky, H.-C. Pauli, S.~S. Pinsky, \emph{Phys.~Rep.}, \textbf{301}:4-6
  (1998), 299--486, arXiv:hep-ph/9705477 and references therein.

\bibitem{NPPF}
V.~A. Franke, {Yu. V. Novozhilov}, S.~A. Paston, E.~V. Prokhvatilov, Focus on
  quantum field theory, chap. Quantization of Field Theory on the Light Front,
  23--81, Nova science publishers, New York, 2005, arXiv:hep-th/0404031.

\bibitem{Wilson}
K.~G. Wilson, T.~S. Walhout, A.~Harindranath, W.-M. Zhang, R.~J. Perry, S.~D.
  Glazek, \emph{Phys.~Rev.~D}, \textbf{49}:12 (1994), 6720--6766,
  arXiv:hep-th/9401153.

\bibitem{Glazek}
S.~D. Glazek, Renormalization of {Hamiltonians} in the Light-Front
  {Fock} Space, 1997, arXiv:hep-th/9706212.

\bibitem{Glazek.Wilson.Phys.Rev.D1993}
S.~D. Glazek, K.~G. Wilson, \emph{Phys.~Rev.~D}, \textbf{48}:8 (1993),
  4214--4218.

\bibitem{Glazek.Wilson.Phys.Rev.D1994}
S.~D. Glazek, K.~G. Wilson, \emph{Phys.~Rev.~D}, \textbf{49}:12 (1994),
  5863--5872.

\bibitem{Glazek.ActaPhys.Polon.B2008}
S.~D. Glazek, \emph{Acta Phys. Polon.~B}, \textbf{39} (2008), 3395--3421,
  arXiv:0810.5258 [hep-th].

\bibitem{Glazek.Phys.Rev.D2001}
S.~D. Glazek, \emph{Phys. Rev.~D}, \textbf{63}:11 (2001), 116006,
  arXiv:hep-th/0012012.

\bibitem{Glazek.Wieckowski.Phys.Rev.D2002}
S.~D. Glazek, M.~Wieckowski, \emph{Phys. Rev.~D}, \textbf{66}:1 (2002), 016001,
  arXiv:hep-th/0204171.

\bibitem{tmf99}
S.~A. Paston, E.~V. Prokhvatilov, V.~A. Franke, \emph{Theor.~Math.~Phys.},
  \textbf{120}:3 (1999), 1164--1181, arXiv:hep-th/0002062.

\bibitem{burlang}
M.~Burkardt, A.~Langnau, \emph{Phys.~Rev.~D}, \textbf{44}:4 (1991), 1187--1197.

\bibitem{tmf97}
V.~A. Franke, S.~A. Paston, \emph{Theor.~Math.~Phys.}, \textbf{112}:3 (1997),
  1117--1130, arXiv:hep-th/9901110.

\bibitem{tmf02}
S.~A. Paston, E.~V. Prokhvatilov, V.~A. Franke, \emph{Theor.~Math.~Phys.},
  \textbf{131}:1 (2002), 516--526, arXiv:hep-th/0302016.

\bibitem{Yad.Fiz.2005}
S.~A. Paston, E.~V. Prokhvatilov, V.~A. Franke, \emph{Phys.~Atom.~Nucl.},
  \textbf{68} (2005), 267--278, arXiv:hep-th/0501186.

\bibitem{Deser.Jackiw.Templeton.Annals.of.Physics.2000}
S.~Deser, R.~Jackiw, S.~Templeton, \emph{Ann. Phys.}, \textbf{281}
  (2000), 409--449.

\bibitem{Deser.Jackiw.Templeton.Phys.Rev.Lett.1982}
R.~Jackiw, S.~Templeton, \emph{Phys.~Rev.~Lett.}, \textbf{48} (1982), 975--978.

\bibitem{Mandelstam}
S.~Mandelstam, \emph{Nucl.~Phys.~B}, \textbf{213} (1983), 149--168.

\bibitem{Leibbrandt}
G.~Leibbrandt, \emph{Phys.~Rev.~D}, \textbf{29} (1984), 1699--1708.

\end{thebibliography}

\end{document}